\documentclass[%
 reprint,
 amsmath,amssymb,
 aps,
]{revtex4-2}

\usepackage{graphicx}
\usepackage{dcolumn}
\usepackage{bm}
\usepackage{mathtools}
\usepackage{color}
\DeclarePairedDelimiter\abs{\lvert}{\rvert}

\newcommand{\nik}[1]{\textcolor{black}{#1}}

\begin{document}

\preprint{APS/123-QED}

\title{Self-organized bistability on \nik{scale-free networks}}

\author{Nikita Frolov$^{1,2}$}
\email{phrolovns@gmail.com}
\author{Alexander Hramov$^{1,2}$}%
 \email{hramovae@gmail.com}
\affiliation{
$^{1}$Center for Neurotechnology and Machine Learning, Immanuel Kant Baltic Federal University, Kaliningrad, Russia\\
$^{2}$Institute of Information Technology, Mathematics and Mechanics, Lobachevsky State University of Nizhny Novgorod, Nizhny Novgorod, Russia
}%

\date{\today}

\begin{abstract}
A dynamical system approaching the first-order transition can exhibit a specific type of critical behavior known as self-organized bistability (SOB). It lies in the fact that the system \nik{can} permanently switch between the coexisting states under the self-tuning of a control parameter. \nik{Many of these systems have a network organization} that should be taken into account to understand the underlying processes in detail. In the present paper, we theoretically explore an extension of the SOB concept on the scale-free network under coupling constraints. As provided by the numerical simulations and mean-field approximation in the thermodynamic limit, SOB \nik{on scale-free networks originates from facilitated criticality reflected} on both macro- and mesoscopic network scales. We establish that the appearance of switches is rooted in spatial self-organization and temporal self-similarity of \nik{the network's critical dynamics} and replicates extreme properties of epileptic seizure recurrences. Our results, thus, \nik{indicate} that the proposed conceptual model is suitable to deepen the understanding of emergent collective behavior behind neurological diseases.
\end{abstract}

\maketitle

\section{Introduction}

\nik{Many natural and man-made systems deliver diverse functionality by} switching between coexisting stable states, or {\it multistability}. In living systems, this phenomenon regulates the processes on different scales: from the interaction of inner organ systems to cell cycles and neuronal synchronization on considerably smaller scales~\cite{angeli2004detection,strogatz2012sync,pisarchik2014control}. Specifically, the ability to switch between the multiple patterns of local coherence on the brain's cortical network underpins many aspects of consciousness~\cite{koch2016neural}. Past works have identified that in normal conditions, emergent neuronal avalanches expose distinct power-law scaling properties, which are the hallmarks of self-organized criticality (SOC).~\cite{linkenkaer2001long,beggs2003neuronal,bak2013nature,shew2013functional}.

\nik{On the contrary,} excessive synchronization involving large-scale networks into coherent motion is an example of undesired and harmful behavior, which is the case of the brain's epileptic condition~\cite{louzada2012suppress,lehnertz2009synchronization,luttjohann2015dynamics}. Such extreme bursts of coherence lie beyond \nik{the power-law} distribution and are most probably rooted in the bistable dynamics of neuronal ensembles. In this regard, Di Santo et al. have extended the theory of SOC and have described these dynamics as a {\it self-organized bistability} (SOB)~\cite{di2016self}. They have proposed a conceptual model of a system in the vicinity of the first-order transition -- a singular point at which both stable states are simultaneously present. The switching between these states is provided by self-tuning the control parameter under the impact of dissipation and driving force. Remarkably, the observed avalanches obey \nik{the statistics of dragon-kings (DKs), which are rare and significantly large events. At the same time, the occurrence of DKs originates from the system's nonlinearity. Therefore these events are not random and even predictable to some extent~\cite{sornette2012dragon}. Overall, these facts confirm} the fundamental significance of the SOB concept in non-trivial neural dynamics under epileptic conditions.

However, to provide further insight into the origins of bistability on damaged neuronal \nik{populations,} one has to account for their networked organization~\cite{kalitzin2019epilepsy}. It is well-known that the interplay between structure and dynamics enriches the available modes of collective behaviors~\cite{boccaletti2006complex,boccaletti2018synchronization}; Thus, it is essential to understand how the structural properties drive SOB on complex networks and give birth to excessively large avalanches. 

In our latest work~\cite{frolov2021extreme}, we have introduced a conceptual network model under coupling constraints that should be considered a network extension of the SOB theory. In this model, the dissipation and driving of the interelement coupling self-tuning, as suggested by SOB, is biologically motivated by the general aspects of neuro-astrocyte interaction. Its dysregulation provides a metabolic mechanism for astrocyte-dependent hyperexcitability~\cite{siracusa2019astrocytes} and is recognized as one of the driving factors of epilepsy~\cite{diaz2019glia,siracusa2019astrocytes}. Another feature of the model that needs to be emphasized is the exploitation of a self-similar {\it scale-free} (SF) structure that determines the bistability of collective behavior, i.e., an explosive transition, under certain conditions~\cite{gomez2011explosive,coutinho2013kuramoto,boccaletti2016explosive}. Although the SF organization of local neuronal populations is debatable and has not been proven so far, there are several indirect signs and arguments based on the underlying criticality of neural dynamics in support of this hypothesis~\cite{shew2013functional,haimovici2013brain,gastner2016topology}. 

Consistent with SOB, our model has replicated the appearance of epileptic seizure generation and self-termination of a supercritical synchronized state due to the interplay between the explosive synchronization and coupling constraints. We referred to such a state as an {\it extreme synchronization event} due to its spontaneous occurrence and short lifetime. Although we have demonstrated that the designed model produces solitary synchronization events, the critical transitions leading to such a hazardous behavior have not been fully addressed.

In this work, we exploit extensive theoretical analysis to derive the conditions of SOB for the proposed network under coupling constraints. \nik{We primarily focus on the} Barab{\'a}si-Albert (BA) model, a paradigmatic model of SF network grown through the preferential attachments~\cite{barabasi1999emergence}. This model yields a degree distribution with constant exponential decay of $-3$. For this model, we derive the mean-field approximation in the thermodynamic limit that is in good agreement with the numerical simulations of a finite-size network. Having analyzed continuous synchronization diagrams and statistical properties of the synchronization events occurrence, we show an essential role of self-organized bistability \nik{in producing and terminating} short-term synchronization. We establish that these events reproduce the statistical properties of extreme value theory and epileptic brain activity. Finally, we generalize our analysis for SF networks with an arbitrary scaling exponent and reveal that such a structural property \nik{has} a particular impact on the emergence of SOB and its temporal self-similarity.

\section{Model}

Consider a network of $N$ Kuramoto phase oscillators, whose rotation is defined by the system of differential equations
\begin{align}
    \frac{d\theta_i}{dt} &= \omega_i + \lambda_i \sum_{j=1}^N \mathcal{A}_{ij} \sin(\theta_j-\theta_i), \label{eq:1}\\
    \frac{d\lambda_i}{dt} &= \alpha(\lambda_0-\lambda_i) -\beta r_i. \label{eq:2}
\end{align}

In Eq.~(\ref{eq:1}), $\theta_i$ and $\omega_i$ denote instantaneous phase and natural frequency of each $i^{th}$ Kuramoto oscillator, $i=1,...,N$. The adjacency matrix $\mathcal{A}_{ij}$ defines the connectivity of the SF network yielding a power-law degree distribution $P(k)\sim k^{-\gamma}$. To construct $\mathcal{A}_{ij}$ we use the BA model that generates a degree-degree correlated SF network based on preferential attachment~\cite{barabasi1999emergence} yielding a degree distribution with fixed scaling exponent $\gamma=3$. In the BA model, $m = 3$ new edges preferentially attach between a new vertex and existing vertices at each step of the growing process. To explore the influence of scaling exponent, we employ the Chung-Lu (CL) that generates an uncorrelated SF network with an arbitrary pre-defined value of scaling exponent $\gamma$~\cite{goh2001universal,chung2002connected}. The CL model is described in detail in Appendix~\ref{sec:appendix_A}. In both models, we adjust $\omega_i=k_i$ to fulfill the frequency-degree correlation. Macroscopic motion of the ensemble is defined by the complex order parameter $Re^{i\Psi}=1/N\sum_{j=1}^Ne^{i\theta_j}$. 

Following Di Santo et al.~\cite{di2016self}, Eq.~(\ref{eq:2}) describes the self-tuning of individual coupling strength $\lambda_i$. Here, coupling consumption (a dissipation) is a function of local order parameter $r_i = 1/k_i \abs{\sum_{j=1}^N \mathcal{A}_{ij}e^{i\theta_j}}$, which implies that the maintenance of a certain level of coherence in the neighborhood of the $i^{th}$ unit consumes its coupling ability at rate $\beta$. Besides, each unit diffusely couples to an external source that provides the recovery of individual coupling ability to the level $\lambda_0$ at rate $\alpha$ (a driving force).

To provide deeper insights into the dynamics of Eq.~(\ref{eq:1})–(\ref{eq:2}) concerning the BA model, we used the mean-field formalism proposed by Ichinomiya~\cite{ichinomiya2004frequency}. Considering the motion of the ensemble at the mean frequency $\Omega = \langle \omega\rangle = \langle k\rangle$ with the average phase $\Psi$, we can replace $\phi = \theta - \Psi$. In a continuous limit, the density of oscillators having phase $\phi$ and degree $k$ at time moment $t$ is defined by $\rho(\phi,k,t)$, such that $\int_0^{2\pi}d\phi\rho(\phi,k,t)=1$, that obeys the continuity equation:
\begin{equation}
    \frac{\partial \rho}{\partial t} + \frac{\partial}{\partial \phi}\left\{v_\phi \rho\right\}=0,
    \label{eq:3}
\end{equation}
Eq.~(\ref{eq:3}) therefore describes the motion of the networked system (\ref{eq:1})-(\ref{eq:2}) in the continuous limit with $v_{\phi}=d\phi/dt$. Let us define the global order parameter through the density $\rho(\phi,k,t)$ in the integral form
\begin{equation}
    R e^{i\Psi} = \frac{1}{\langle k\rangle}\int_m^{k_{max}}dk\int_{0}^{2\pi} d\phi kP(k)\rho e^{i\phi},
    \label{eq:4}
\end{equation}
and rewrite Eq.~(\ref{eq:1}) using the obtained definition (\ref{eq:4}) to find the mean-field approximation of $v_\phi$:
\begin{equation}
    v_\phi=\frac{d\phi}{dt}=(1-\lambda R \sin\phi)k - \langle k\rangle.
    \label{eq:5}
\end{equation}

In Eq.~(\ref{eq:5}), we suggest that $\lambda_i=\lambda$ and $r_i=R$. Given (\ref{eq:4}) and (\ref{eq:5}), Peron and Rodrigues found the steady-state solution of Eq.~(\ref{eq:3}) for a frequency-degree correlated SF network~\cite{peron2012determination}:
\begin{equation}
    \rho(k,\phi) =
    \begin{cases}
    \delta\left(\phi-\arcsin\left[\frac{k-\langle k\rangle}{\lambda Rk}\right]\right), \quad \abs{k-\langle k\rangle}\leq \lambda Rk\\
    \frac{A(k)}{\abs{(k-\langle k\rangle)-\lambda Rk \sin\phi}}, \quad \text{otherwise},
    \end{cases}
\end{equation}
with normalization term $A(k)$.

The condition $\abs{k-\langle k\rangle}\leq \lambda Rk$ is fulfilled for the phase-locked fraction of networked oscillators while remaining possess incoherent drift. Since only phase-locked units contribute to the global coherence of the ensemble, Eq.~(\ref{eq:4}) becomes
\begin{equation}
    R=\frac{1}{\langle k\rangle}\int_{\abs{k-\langle k\rangle}\leq \lambda Rk} kP(k)\sqrt{1-\left(\frac{k-\langle k\rangle}{\lambda Rk}\right)^2}dk.
    \label{eq:7}
\end{equation}

To complete a mean-field approximation of the system (\ref{eq:1})-(\ref{eq:2}) we should complement Eq.~(\ref{eq:7}) with account of coupling constraints. In the steady state $d\lambda/dt=0$, it becomes
\begin{equation}
    \lambda = \lambda_0 - \frac{\beta}{\alpha}R.
    \label{eq:8}
\end{equation}

To provide rigorous conclusions, we present Eqs. (\ref{eq:7})-(\ref{eq:8}) in the thermodynamic limit $N\rightarrow\infty$ by explicitly taking the exact degree distribution $P^{\infty}(k)=2m(m+1)/(k(k+1)(k+2))$ and mean degree $\langle k\rangle^{\infty}=2m$ for BA model ~\cite{dorogovtsev2000structure,krapivsky2000connectivity}. Further, we denote a mean-field approximation of the order parameter as $R^\infty$.

\nik{Using mean-field analysis one can also} estimate conditions for the trivial solution of Eq.~(\ref{eq:7}), $R^\infty = 0$, to lose its stability. \nik{By letting $R^\infty\rightarrow+0$,} Peron and Rodrigues derived this critical point, which in thermodynamic limit is $\lambda_c=1/(\pi m P^\infty(2m))\simeq1.485$~\cite{peron2012determination}. Accounting for coupling constraints we define the critical point for our model (\ref{eq:1})-(\ref{eq:2}) as:
\begin{equation}
    \lambda_{0c}=\frac{1}{\pi m P^\infty(2m)} + \frac{\beta}{\alpha}R^0,
    \label{eq:9}
\end{equation}
where $R^0=1/\sqrt{N}$ is a finite estimation of the order parameter in the incoherent state.

\begin{figure}[!t]
	\begin{center}
		\includegraphics[width=8.6 cm]{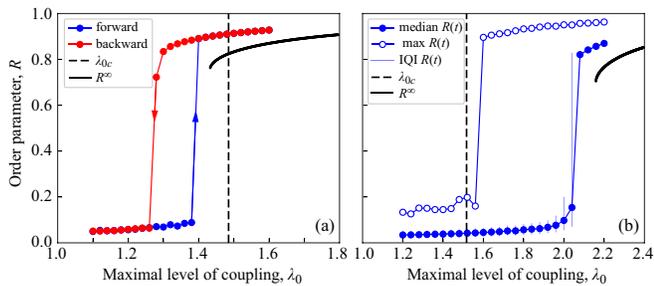}
	\end{center}
	\caption{(a) (a) Forward and backward synchronization diagrams without excitability constraints $\beta=0$. (b) A forward diagram in the presence of a finite excitability consumption $\beta=0.04$. Synchronization diagrams in both panels are computed for fixed $\alpha=0.04$. In panel (b), the median value, the maximum value, and the inter-quartile interval (IQI) of $R(t)$ were computed over the time series of $2\times 10^4$ time units. In both panels, the black solid line and vertical dashed line show a mean-field approximation $R^{\infty}$ and the critical point $\lambda_{0c}$, respectively.}
	\label{fig:1}
\end{figure}

\section{Results}

We start with considering forward and backward continuations of a described network by increasing and decreasing $\lambda_0$ using increment $\delta \lambda_0=0.02$. Eqs.~(\ref{eq:1})-(\ref{eq:2}) for a finite-size network are integrated for $N=1000$ unit.

Fig.~\ref{fig:1} reports synchronization diagrams $R(\lambda_0)$ in the absence and presence of coupling constraint for fixed $\alpha=0.04$. As expected, without consumption $\beta=0$ (Fig.~\ref{fig:1}a), our system exhibits an explosive transition reproducing a hysteresis loop demonstrated in the pioneering work on a frequency-degree correlated SF network by G{\'o}mez-Garde{\~n}ez et al.~\cite{gomez2011explosive}. A mean-field approximation also predicts an explosive transition indicating a bistability region, where both $R^\infty=0$ and $R^\infty\neq0$ are stable ($\lambda_{0c}\simeq1.485$).

Numerical results for a system under coupling consumption $\beta=0.04$ show that the onset of forward transition is largely delayed concerning the critical point $\lambda_{0c}$. Fig.~\ref{fig:1}b displays that both coherent and incoherent solutions are possible in the region between $\lambda_{0c}$ and the onset of forward transition. It follows from Eq.~(\ref{eq:2}) that the stable coherent state is only possible if coupling consumption is compensated by its recovery $\alpha (\lambda_0-\lambda_i)\geq \beta r_i$ for each network unit. Since the latter condition is not fulfilled in the area between $\lambda_{0c}$ and the onset of forward transition, the turbulent state $R\simeq 0$ here remains preferable for the network. However, rare transitions to synchronized state shown in Fig.~\ref{fig:1}b) as the outliers of $R(t)$ distribution are possible due to $\lambda_0>\lambda_{0c}$. 

\begin{figure}[!t]
	\begin{center}
		\includegraphics[width=8.6cm]{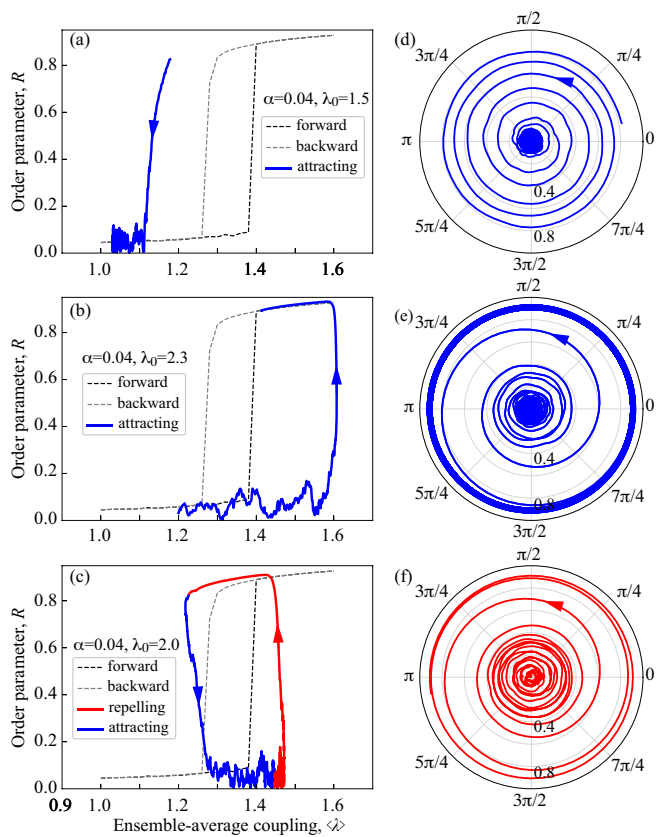}
	\end{center}
	\caption{(a)-(c) Phase portraits of the macroscopic dynamics of the SF network under coupling constraints on the plane $(R,\langle \lambda\rangle)$ for fixed values of $\beta=0.04$, $\alpha=0.04$ and different $\lambda_0$: (a) $\lambda_0=1.5$ -- subcritical dynamics; (b) $\lambda_0=2.3$ -- supercritical dynamics; (c) $\lambda_0=2.0$ -- critical dynamics. \nik{Dashed black and gray lines in (a)-(c) display the forward and backward synchronization diagrams $R(\lambda)$ for $\beta=0$.} Corresponding phase portraits on the complex plane in the polar coordinate system $(R,\Psi)$ are presented in (d)-(f).}
	\label{fig:2}
\end{figure}

Based on Fig.~\ref{fig:1}b, three characteristic areas can be distinguished: (i) the region of subcritical dynamics $\lambda_0<\lambda_{0c}$; (ii) the region of critical bistable dynamics between $\lambda_{0c}$ and the onset of forward transition; (iii) the region of supercritical dynamics after forward transition corresponding to a coherent motion.

Now, we illustrate the network's dynamics in terms of its macroscopic parameters under variation of $\lambda_0$. Fig.~\ref{fig:2} reports macroscopic motion on the parameter plane $(R,\langle \lambda\rangle)$, where $\langle \lambda\rangle = 1/N\sum_{i=1}^N \lambda_i$ is the ensemble-average level of coupling ability, accompanied by corresponding trajectories on the complex polar plane $(R,\Psi)$. As expected, in a subcritical region, the system converges to a turbulent drift around the fixed point at a lower branch of the hysteresis curve (Fig.~\ref{fig:2}a,d). In a supercritical mode, it is attracted to an upper branch that corresponds to the limit cycle at mean frequency $\Omega=\langle k\rangle$ (Fig.~\ref{fig:2}b,e).

\begin{figure}[!t]
	\begin{center}
		\includegraphics[width=8.6 cm]{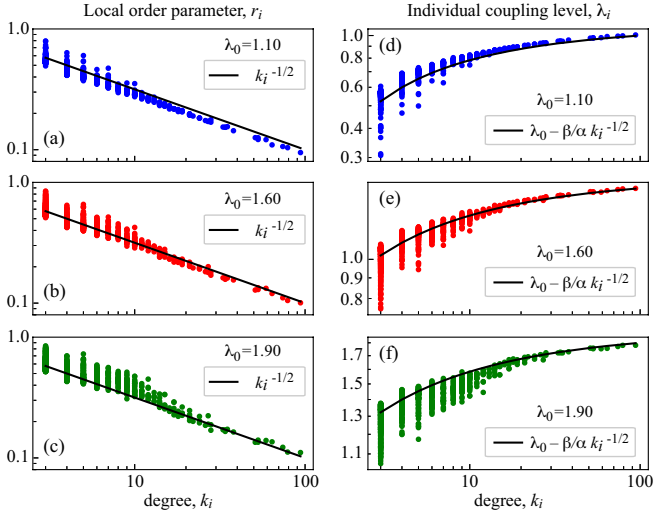}
	\end{center}
	\caption{Distribution of mesoscopic parameters -- local order parameter $r_i$ (a)-(c) and individual coupling level $\lambda_i$ (d)-(e) -- versus node's degree $k_i$. The distributions are presented for fixed $\alpha=0.04$, $\beta=0.04$ and different values of $\lambda_0$: (a),(d) $\lambda_0=1.1$ -- subcritical behavior; (b),(e) $\lambda_0=1.6$ -- weak criticality; (c),(f) $\lambda_0=1.9$ -- strong criticality. Black lines indicate approximations $r_i=1/\sqrt{k_i}$ in left column and $\lambda_i=\lambda_0-\beta/(\alpha\sqrt{k_i})$ in right column.}
	\label{fig:3}
\end{figure}

\nik{Remarkably, in the region of critical bistable dynamics, the system mostly drifts near the fixed point $R\simeq0$ in the vicinity of transition point $\lambda_c\simeq1.485$. However, the turbulent drift may occasionally push the trajectory beyond the critical point. Hence, the network explosively synchronizes as reflected by motion along the repelling trajectory toward the upper branch (Fig.~\ref{fig:2}c). Having landed the upper branch the network moves to backward transition forced by resource consumtion. As the network exhausts the coupling resource, it undergoes the backward transition. Eventually, the diffusive mechanism of coupling recovery returns the trajectory to the starting position near the forward transition point on the lower branch of the hysteresis curve. The reported process repeats over time as the coupling becomes excessively strong $\langle \lambda\rangle>\lambda_{c}$ due to chaotic fluctuations in the vicinity of the critical point.}

The criticality of the considered Kuramoto model under coupling constraints that determines the spontaneous appearance of synchronization events is reflected on the mesoscopic level, i.e., on the level of local populations. Log-log plots in Fig.~\ref{fig:3} show the distributions of local order parameter $r_i(k_i)$ and individual coupling ability $\lambda_i(k_i)$ during a long-term period of turbulent motion ($2\times 10^4$ time units). In a subcritical regime $\lambda_0=1.1$, local order parameter is well-approximated by inverse square root law $r_i=1/\sqrt{k_i}$ (Fig.~\ref{fig:3}a). It indicates that below the critical value $\lambda_{0c}=1.517$, phases within the local groups are homogeneously distributed on the unit circle, implying turbulent motion across all the mesoscopic scales. Substituting $r_i=1/\sqrt{k_i}$ in Eq.~(\ref{eq:2}) and letting $d\lambda_i/dt=0$, one finds the stationary distribution of individual coupling ability in a subcritical regime $\lambda_i=\lambda_0-\beta/(\alpha\sqrt{k_i})$ that perfectly fits the empirical distribution at $\lambda_0=1.1$ (Fig.~\ref{fig:3}d).

\begin{figure}[!t]
	\begin{center}
		\includegraphics[width=8.6 cm]{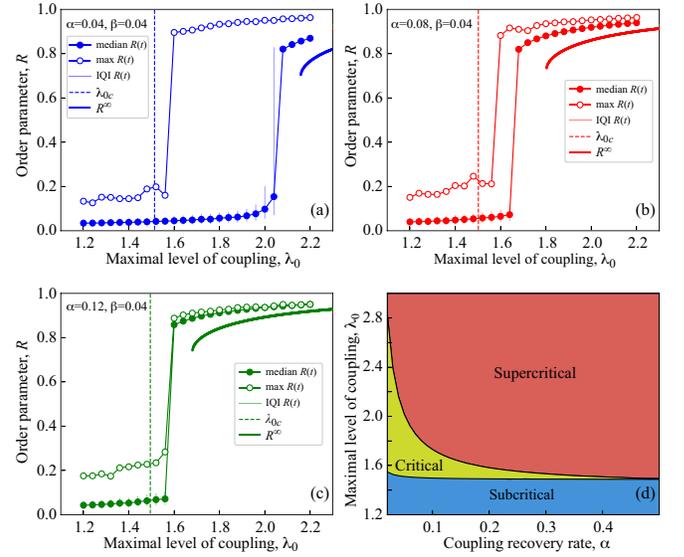}
	\end{center}
	\caption{Synchronization diagrams $R(\lambda_0)$ for the fixed value $\beta=0.04$ and different $\alpha$: (a) $\alpha=0.04$; (b) $\alpha=0.08$; (c) $\alpha=0.12$. (d) Phase diagram in the $(\alpha,\lambda_0)$ parameter plane plotted upon the mean-field approximation given by \nik{Eqs.~(\ref{eq:7})--(\ref{eq:8})}.}
	\label{fig:4}
\end{figure}

At $\lambda_0=1.6$ slightly exceeding the critical value $\lambda_{0c}$ (weak criticality, Fig.~\ref{fig:3}b,e), inverse square root approximations $r_i,\lambda_i\sim1/\sqrt{k_i}$ are applicable only to large-scale populations formed around \nik{the network's hubs}. At the same time, the local coherence of peripheral groups around low- and medium-degree units starts rising above the estimate given by this approximation. \nik{With increasing criticality (Fig.~\ref{fig:3},c,f), local coherence around the low- and medium-degree nodes grows while hubs remain desynchronized.} Taking into account that $R\simeq1/\sqrt{N}$ in all considered cases, facilitation of local coherence does not contribute to global synchronization. Instead, the critical dynamics of this model form a complex of populations synchronized on different spatial and temporal scales and weakly interacting with each other. Noteworthy, due to the coupling consumption, the coherence of small groups is not long-term preserved but increases and decreases over time independently of other groups, thus resisting a transition to global synchronization. However, as soon as a large number of local populations with similar frequencies accidentally synchronize, at the same time, critical dynamics create conditions for synchronizing key components -- \nik{the network's hubs}. Involvement of hubs in the coherent motion launches an avalanche-like process of a large-scale phase-locking resulting in an abrupt first-order transition to metastable global synchrony.

\begin{figure}[!t]
	\begin{center}
		\includegraphics[width=8.6 cm]{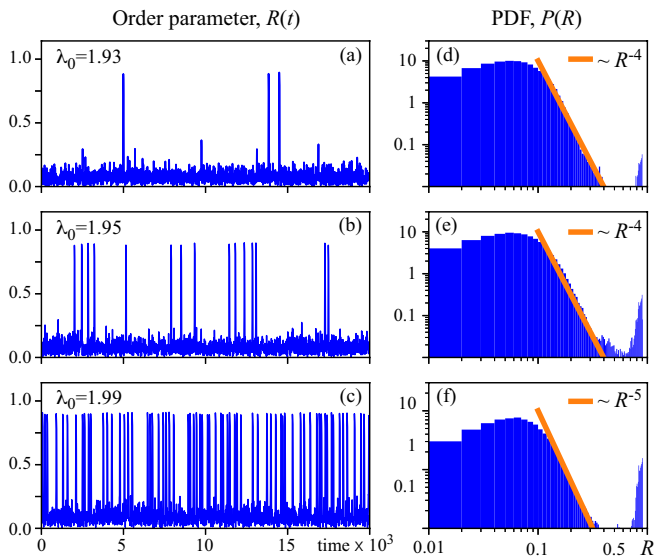}
	\end{center}
	\caption{Long-term time series $R(t)$ (a)-(c) and corresponding log-log histograms (d)-(f) in the proximity of forward transition for fixed $\alpha=0.04$, $\beta=0.04$ and different $\lambda_0$: (a),(d) $\lambda_0=1.93$; (b),(e) $\lambda_0=1.95$; (c),(f) $\lambda_0=1.99$. The orange line in (d)-(f) shows a fitted power law.}
	\label{fig:5}
\end{figure}

\begin{figure*}[!t]
	\begin{center}
		\includegraphics[width=15 cm]{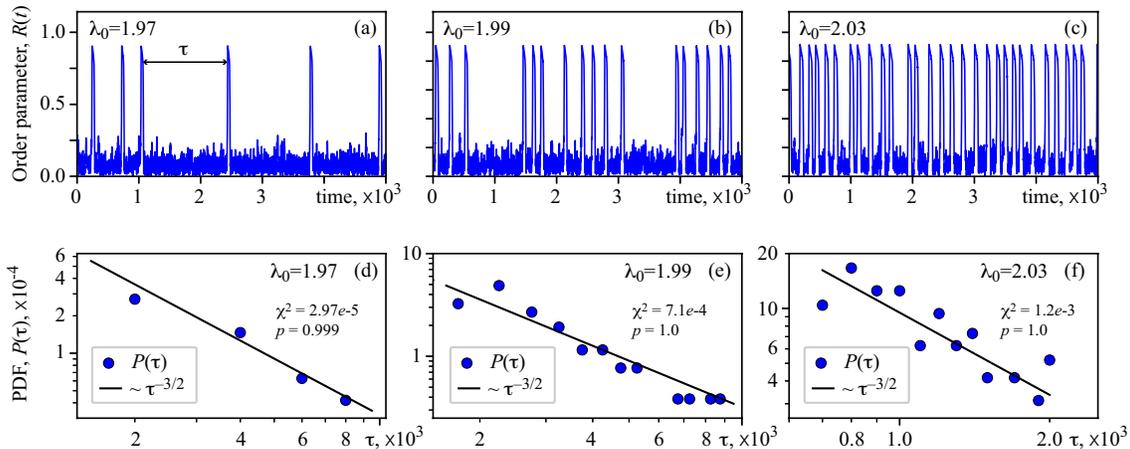}
	\end{center}
	\caption{Long-term time series $R(t)$ (a)-(c) and corresponding distributions of return intervals (d)-(f) in the proximity of forward transition for fixed $\alpha=0.04$, $\beta=0.04$ and different $\lambda_0$: (a),(d) $\lambda_0=1.97$; (b),(e) $\lambda_0=1.99$; (c),(f) $\lambda_0=2.03$. Black lines in (d)-(f) show a fitted power law $\sim\tau^{-3/2}$. The outcomes of the $\chi^2$-test are presented within each panel. }
	\label{fig:6}
\end{figure*}

In addition, we demonstrate how the balance between recovery rate $\alpha$ and consumption rate $\beta$ affects the critical dynamics of the considered model. Fig.~\ref{fig:4} reports the synchronization diagrams $R(\lambda_0)$ for different values of $\alpha$ at fixed $\beta=0.04$. It follows from \nik{Eqs.~(\ref{eq:7})--(\ref{eq:8})} that increasing $\alpha$, or equivalently decreasing $\beta$, reduces the impact of coupling constraints on the system's dynamics. Indeed, the region of critical behavior bounded by the critical value $\lambda_{0c}$ from the left and the onset of forward transition from the right shrinks with increasing recovery rate $\alpha$ (Fig.~\ref{fig:4}a,b) until the transition to coherence becomes of a first-order (Fig.~\ref{fig:4}c). This observation for a finite-size network is consistent with the mean-field approximation in the thermodynamic limit $N\rightarrow\infty$ (Fig.~\ref{fig:4}d).

Up to now, we have considered the emergence of such synchronization events and associated critical dynamics of the Kuramoto network under coupling constraints. Now, we explore the statistics of their appearance in the proximity of forward transition in terms of {\it extreme events theory}. Let us fix $\alpha=0.04$, $\beta=0.04$ and consider the Kuramoto ensemble under coupling constraints approaching the point of forward transition. Although the area of criticality is broad enough, as displayed in Fig.~\ref{fig:1}b and Fig.~\ref{fig:4}a, the generation of solitary synchronization events is quite rare and occasional in a long-term perspective, not exceeding one event per $2\times10^4$ time units. However, the probability of their emergence drastically increases in the proximity of forward transition point $\lambda_0\simeq2.05$ (Fig.~\ref{fig:5}), allowing us to explore the statistical properties of these events.

Fig.~\ref{fig:5}a-c show the time series $R(t)$ and corresponding histograms (Fig.~\ref{fig:5}d-f) with increasing $\lambda_0$. As $\lambda_0$ approaches the forward transition point, the number of produced metastable coherent events increases. \nik{Analysis of the histograms reveals that the variation of $R$ near the preferable incoherent state obeys power law \nik{$P(R)\sim R^{-\zeta}$}. Although the power law is fitted in a relatively narrow region of $R$ values near the stable fixed point, this observation could be considered a hallmark of self-organized criticality~\cite{bak2013nature}.} It implies that the occurrence of small- and medium-size events, i.e., the short-term establishment of coherence in small- and medium-size groups, obeys scaling of the same physical mechanism. Although the dynamics seem to be highly turbulent at first glance, the power-law fit of the tail indicates the presence of spatial order. At the same time, a peak beyond the power-law fit indicates a non-negligible probability of large deviations of $R$. Such events lie outside the scaling rule and obey different physical principles. Such distribution is a hallmark of a specific type of extreme events -- dragon kings, DKs~\cite{sornette2012dragon,mishra2018dragon}. DKs possess a remarkable property -- these events are significant and non-random; therefore, they are predictable to some degree. Indeed, DKs are generated in a deterministic system and obey certain mechanisms underlying critical behavior in the vicinity of the typing point. The possibility of predicting such states, although tangible, is a challenging task that requires a deep understanding of the structure and dynamics of the considered networked system.

\begin{figure}[!t]
	\begin{center}
		\includegraphics[width=8.6 cm]{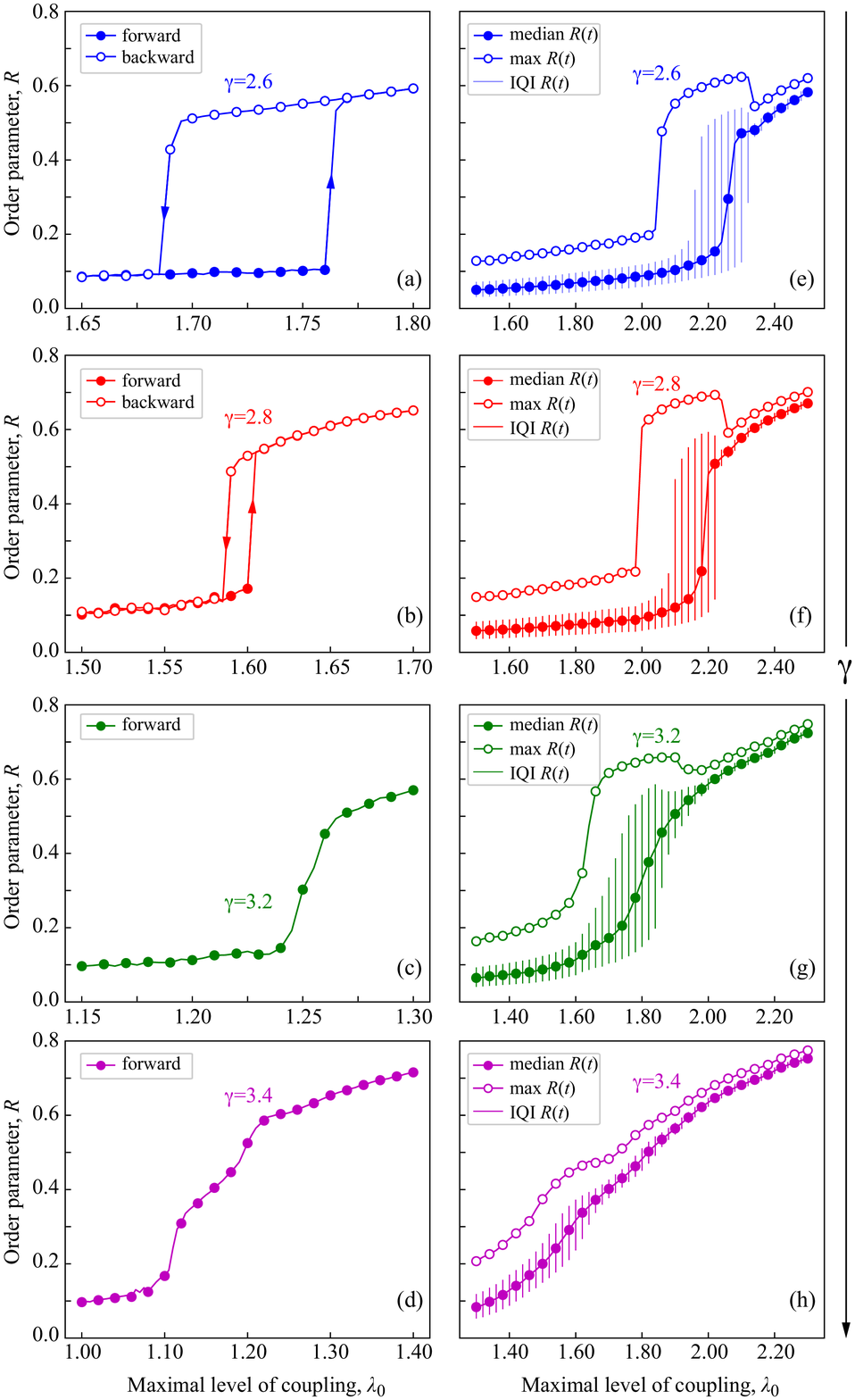}
	\end{center}
	\caption{Forward and backward synchronization diagrams in the absence of excitability constraints $\beta=0$ for different values of $\gamma$: (a) $\gamma=2.6$, (b) $\gamma=2.8$, (c) $\gamma=3.2$, and (d) $\gamma=3.4$. (e)-(h) Respective forward diagrams in the presence of a finite excitability consumption $\beta=0.04$. Synchronization diagrams in both panels are computed for fixed $\alpha=0.04$. In panels (e)-(h), the median value, the maximum value, and the inter-quartile interval (IQI) of $R(t)$ were computed over the time series of $2\times 10^4$ time units.}
	\label{fig:7}
\end{figure}

To finalize our analysis of the BA model, we explored the distribution of return intervals between neighboring synchronization events $\tau$. Fig.~\ref{fig:6} displays long-term time series $R(t)$ and corresponding return intervals distributions $P(\tau)$ for different $\lambda_0$ in the proximity of forward transition. Despite an increase of criticality and changes in characteristic scales of return intervals at different $\lambda_0$, their distribution is well-fitted by the power law $P(\tau)\sim\tau^{-3/2}$. The Pearson's $\chi^2$-test, which has been used to assess the goodness of fit, yields a correspondence between the observed and expected power-law distributions (test outcomes are presented in each subplot). Similar to the previous discussion on the histograms $P(R)$, a distinct power-law fitting of $P(\tau)$ evidences the presence of temporal order in the sequence of synchronization events despite being aperiodic and visually irregular. This observation is most probably rooted in the concept of ``on-off'' intermittency -- an alteration of turbulent (``on'') and laminar (``off'') phases of the system's motion in the vicinity of synchronization onset~\cite{hramov2006ring}. In ``on-off'' intermittency, an analogous power-law fit with characteristic exponent $-3/2$ defines the distribution of laminar phases duration and reflects the approach to ordered dynamics from turbulence as the system evolves towards synchronization. The $-3/2$ power law is also reported in empirical observations of epileptic activity in the rodents' brain~\cite{hramov2006off,sitnikova2012off,koronovskii2016coexistence,frolov2019statistical}. For instance, Refs.~\cite{hramov2006off,sitnikova2012off,koronovskii2016coexistence} propose a theoretical framework bonding ``on-off'' intermittency and occurrence of epileptic seizures. Ref.~\cite{frolov2019statistical} explicitly demonstrates both extreme properties of seizures' amplitudes and temporal scaling fitted by the $-3/2$ power law.

\nik{Lastly, we investigate how the structure of the SF network affects SOB.} Specifically, we are interested in the influence of scaling exponent $\gamma$ since this parameter regulates the prevalence of highly-connected hubs in SF networks. The latter units play a principal role in the first-order transition and, consequently, the emergence of an SOB in our model. Earlier, Coutinho et al.~\cite{coutinho2013kuramoto} have developed a theoretical framework to show that in the frequency-degree correlated networks, the first-order transition is present at $2<\gamma<3$. Recall that this particular range of scaling exponent is the most common in real SF networks~\cite{barabasi2013network}. For $\gamma>3$, Coutinho et al. have predicted the suppression of hysteresis that determines the continuous type of transition.

Using the CL model, we have generated the degree-degree uncorrelated SF networks with $\gamma = \{2.6, 2.8, 3.2, 3.4\}$. Fig.~\ref{fig:7} displays the synchronization diagrams $R(\lambda_0)$ for these networks in the absence and presence of coupling constraint for fixed $\alpha=0.04$. Consistent with the work by Coutinho et al.~\cite{coutinho2013kuramoto}, our results confirm that in the absence of coupling constraints, the network undergoes an explosive transition at $\gamma<3$ and a continuous transition at $\gamma>3$ (Fig.~\ref{fig:7}a-d). As in~\cite{coutinho2013kuramoto}, one can also see that the hysteresis area shrinks as $\gamma$ increases to $3$. Introduction of coupling constraints at rate $\beta=0.04$ results in a distinct SOB behavior for $\gamma<3$. In Fig.~\ref{fig:7}e,f, one can see the area of critical dynamics followed by an abrupt stabilization of a globally coherent state, i.e., a discontinuous transition of median $R(t)$. With a further increase of $\gamma$, the area of critical dynamics associated with SOB vanishes as the transition to coherence becomes continuous (Fig.~\ref{fig:7}g,h). Actually, for $\gamma>3$, the very concept of SOB itself is meaningless since the network is \textit{no longer bistable}. However, there are still at least two reasons for the SF network to exhibit SOB-like behavior at $\gamma=3.2$ as displayed by Fig.~\ref{fig:7}g. First, it could be due to a consideration of a finite-sized and relatively small network ($N=1000$). In this case, the scaling of $\gamma=3.2$ could still provide a sufficient amount of highly-connected units for the system to synchronize rapidly yet continuously. Second, it could be due to the chosen values of $\alpha$ and $\beta$, i.e., coupling recovery and consumption rates. With $\alpha=\beta=0.04$, a continuous transition runs much faster than the self-tuning of coupling strength $\lambda$ that determines sufficiently rapid but \textit{not bistable} switches.

\begin{figure}[!t]
	\begin{center}
		\includegraphics[width=7.6 cm]{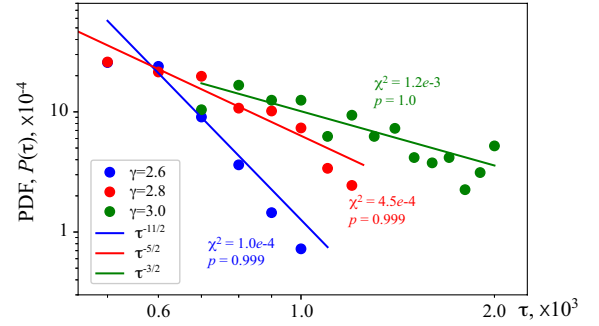}
	\end{center}
	\caption{Distributions of return intervals in the proximity of forward transition for fixed $\alpha=0.04$, $\beta=0.04$ and different scaling exponents $\gamma$ in the CL model: $\gamma=2.6$, $\lambda_0=2.16$ (blue dots) and $\gamma=2.8$, $\lambda_0=2.06$ (red dots). Green dots repeat the distribution for the BA model with $\gamma=3.0$ and $\lambda_0=2.03$. Solid lines of respective colors show the power-law fit. The outcomes of the $\chi^2$-test are presented for each line.}
	\label{fig:8}
\end{figure}

Besides the above-discussed effect, the variation of scaling exponent $\gamma$ within the SOB-associated range $2<\gamma<3$ impacts the slope of distribution of return intervals. Fig.~\ref{fig:8} displays such distributions for $\gamma$ equal to $2.6$, $2.8$ and $3.0$ along with the respective power-law fits. One can see that decreasing SF scaling exponent $\gamma$ increases the slope of return intervals distribution. Although the particular reasons for these observations require further investigation and are yet to be discovered, one may conclude that the SF network's structure is closely related to the properties of its temporal self-similarity. Indeed, increasing the amount of highly-connected hubs on the SF network might cause more frequent bistable switches while maintaining the distribution's exponential slope. At the same time, the slope of return intervals distribution converges to the universal constant $-3/2$ as $\gamma\rightarrow 3$ since $\gamma=3$ is the point of critical singularity, separating the first- and second-order transitions on the frequency-degree correlated SF networks.

\section{Conclusion}

To conclude, this paper reports a theoretical study of the self-organized bistability of a scale-free Kuramoto network under coupling constraints. Our analysis shows that the imbalance between the rates of coupling consumption and recovery determines the delay between the forward synchronization and the critical point, at which the global incoherence becomes unstable. Facilitated critical dynamics of the networked ensemble in this area provide self-consistent switching between coexisting states of incoherence and coherence. Current observations demonstrate that the appearance of such switches bistable exhibits the pronounced statistical properties of extreme events and reproduces the features of epileptic seizures recurrences. It evinces a high degree of self-similarity and ordered dynamics behind the process looking uncorrelated at first glance. Finally, we establish that emergence of self-organized bistability on a scale-free is determined by its structural properties, e.g., in terms of its scaling exponent $\gamma$. We reveal that this type of critical behavior is observed at $2<\gamma<3$, which is the most common in real scale-free networks and suppressed at $\gamma>3$. Our conceptual and yet biologically motivated network model sheds light on the macro- and mesoscopic properties of collective behavior behind the disruptive emergent hypersynchronization of the brain networks paving the way for their study in dynamical and graph-theoretical contexts.

\begin{figure*}[!t]
	\begin{center}
		\includegraphics[width=16 cm]{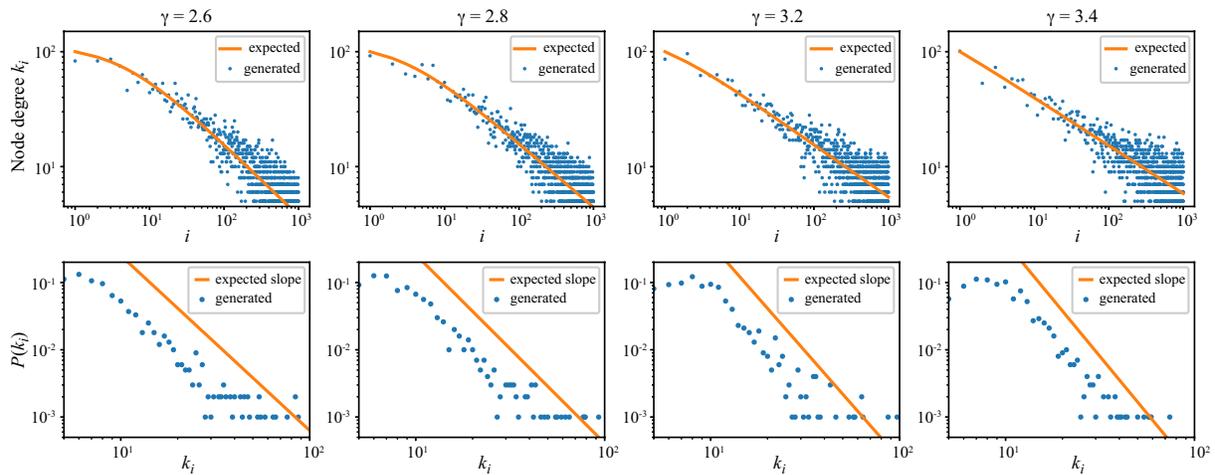}
	\end{center}
	\caption{Constructed uncorrelated scale-free networks according to the Chung-Lu model. Top row presents  expected degree profiles $k'_i (i)$ (orange curves, determined by Eq.~(\ref{eq:exp_deg_profile})) and actual degree profiles of generated scale-free networks $k_i(i)$ (blue dots) for different values of scaling exponent $\gamma$. The bottom row presents respective degree distributions $P(k)$ of generated uncorrelated networks (blue dots) along with the expected slope $\sim k^{-\gamma}$ (orange curves).}
	\label{fig:app}
\end{figure*}

\begin{acknowledgments}
The work has been supported by the Russian Science Foundation (Grant No. 21-72-10129).
\end{acknowledgments}

\appendix

\section{Generating a scale-free graph with arbitrary exponent}
\label{sec:appendix_A}

The Chung-Lu model has been used to produce an uncorrelated scale-free graph with arbitrary scaling exponent~\cite{goh2001universal,chung2002connected}. Specifically, we exploited a recent algorithm proposed by Fasino et al.~\cite{fasino2021generating}. Briefly, it generates a random graph with an arbitrary power-law degree distribution $P(k)\sim k^{-\gamma}$ by adjusting a specific degree to each $i^{th}$ unit according to the expected degree profile:

\begin{equation}
    k'_{i}=c(i_0+i)^{-p}, \qquad p=\frac{1}{1-\gamma}.
    \label{eq:exp_deg_profile}
\end{equation}

In Eq.~(\ref{eq:exp_deg_profile}), parameters $c$ and $i_0$ are defined as:

\begin{equation}
    c=(1-p)k^{*}N^p, \quad i_0=\left(\frac{c}{k_{max}}\right)^{1/p}-1.
\end{equation}

Here, $k^*$ and $k_{max}$ are mean and maximal expected degree. For our calculations, we have set $k^*=10$ and $k_{max}=100$. Degree distributions of constructed scale-free networks for different values of scaling exponent $\gamma$ are presented in Fig.~\ref{fig:app}.

\nocite{*}

\newpage
%

\end{document}